# NectaRSS, an RSS feed ranking system that implicitly learns user preferences

*Juan J. Samper, Pedro A. Castillo, Lourdes Araujo, J. J. Merelo*

## ABSTRACT

In this paper a new RSS feed ranking method called *NectaRSS* is introduced. The system recommends information to a user based on his/her past choices. User preferences are automatically acquired, avoiding explicit feedback, and ranking is based on those preferences distilled to a user profile. NectaRSS uses the well-known vector space model for user profiles and new documents, and compares them using information-retrieval techniques, but introduces a novel method for user profile creation and adaptation from users' past choices. The efficiency of the proposed method has been tested by embedding it into an intelligent aggregator (RSS feed reader), which has been used by different and heterogeneous users. Besides, this paper proves that the ranking of newsitems yielded by NectaRSS improves its quality with user's choices, and its superiority over other algorithms that use a different information representation method.

## *1. Introduction*

The blogosphere offers millions of weblogs on different topics and in different languages. Daily browsing even of a small percentage of these weblogs can be very tedious and unapproachable in practice. RSS feed aggregators, which read RSS feeds chosen by the user to a desktop program, or to a website, avoid website-to-website browsing, but even so, the task of selecting what to read from a few dozen feeds (which include not only weblogs, but also mainstream media sites, and even website updates from sites such as arXiv[1]) usually exceeds practical limits. Users often get tired of checking information even before reaching whatever they are interested in.

In this paper, we propose the **NectaRSS** system, for filtering information gathered from the Web by scoring it according to the user implicit preferences, that is, preferences obtained with the only effort of clicking in whatever newsitem he/she is going to actually read. The system incrementally builds user profiles based on the content (heading or extended content) of these choices.

These techniques will be applied in a novel way to an aggregator of contents to endow it with a certain degree of "intelligence", by ordering the information recovered according to the user profile. Experiments have shown that the results of NectaRSS largely improve those obtained offering the information sorted at random and also using a simple binary algorithm for which relevant documents are those containing the query.

This paper is organized as follows: in Section 2, we review work focused on personalized search systems. In Section 3, we propose novel approaches to providing relevant information that satisfies each user's information need by capturing changes in the user's preferences without the user's effort. In Section 4, we present the experimental results for evaluating our proposed approaches. Finally, we conclude the paper with a summary and directions for future work in Section 5.

---

[1] *http://arXiv.org. the site for Physics (and other disciplines too) preprints.*

## 2. State of the art

**Recommendation systems** have quickly evolved within interactive Web environments. Along this line, Schafer [10] establishes a taxonomy of recommendation systems attending to three categories of features: income and exit functionalities, recommendation methods and design dependent aspects. Middleton [4] presents the recommendation system *Quickstep* to find scientific and research papers. The user preferences are acquired by monitoring his/her behaviour when navigating on the Web, applying automatic learning techniques associated with an ontological representation. Mizzaro [5] uses personalization techniques to implement systems to access electronic publications. It is done by distinguishing between persistent personalization and ephemeral personalization, and it is applied to filtering and retrieval information systems through a specialized web portal.

Merelo [3] proposes a system to recommend to a weblog reader other weblogs on related topics. The system uses the results of a pool and applies association rules. The goal is to find *attribute - value* conditions which appear frequently in a data set. This system considers a set of attributes composed by the URLS appearing in the weblogs and a pools database which indicates if a user has read or not each weblog.

These recommendation systems have not been so far applied to content aggregators, which are a relatively recent product. A content aggregator collects information distributed in different formats, such as **RSS**[2] or **Atom**[3], and periodically checks the updating of its information in order to inform the user. There is a large list of aggregator programs[4], most of which offer similar functionalities. In this paper we present the first recommendation system that works as an aggregator and ranks news items according to a user profile that is automatically computed from past user choices.

## 3. NectaRSS

The system that we propose, designated *NectaRSS*, is designed to rank newly arrived information according to an automatically elaborated user profile. We will restrict our system to information that appears periodically and whose structure is similar to a news story. Thus, the information the system retrieves will be generically referred to as a **newsitem**, which will be composed by a **headline**, a **hyperlink** to their content and optionally a summary. Information aggregators usually show the headline, with a link to the content, and some times the summary; besides, the hyperlink is a unique ID for the newsitem. We will assume that if a newsitem is shown in the aggregator, and the user clicks on it, it corresponds to a topic in which the user is interested, and thus, it will be used to build his/her profile.

Furthermore, NectaRSS uses **sessions**; each *session* is a complete execution of the system, understood as the recovery and scoring of the information available on the Web in that particular moment, according to the preferred sources, the monitoring of user choices and the calculation of the user profile at the end of the execution of the system. This ensemble of techniques used in each session is original and exclusive to the system NectaRSS, configuring a kind of "intelligent aggregator of contents".

---

[2] *RSS is acronym of "Really Simple Syndication".*
[3] *Atom it is other technology to distribute and update contents.*
[4] *RSSfeeds. The RSS, Atom and XML directory and resource, March 2006. On line:*
 *http://www.rssfeeds.com/readers.php*

## 3.1. User Profile Construction Based on Browsing History

In our approach, the user profile is built in an implicit way: the user will not have to take any additional actions such as explicit feedbacks or evaluations to build his or her profile, which will be constructed automatically according to his/her navigation history by the news headlines which are presented to him/her.

A user profile $P$ will be developed during many sessions with the system and kept for use in future sessions. Information about user actions will be gathered in each session and into a session profile $P_s$, which will be incorporated into the user profile at the end of the session. A user can accomplish different sessions during a day and he/she will select different headlines during these sessions. In our method, we will assume that the preferences of the user are built by accumulation of their past selections. In this way, we incrementally build the user profile $P$ considering the cumulated preferences stored in $P$ and the preferences of each session stored in $P_s$. Thus, $P$ will reflect a user profile built with the history of navigation on news during $S$ sessions.

Newsitems and the user profile will be represented using the vector space model proposed by Salton [6, 7]. Thus, we define $S_j$ ($j = 1, 2, ..., N$) as the number of headlines that the user has chosen in session $j$. In each session, $P_s$ will be built through the following process. First of all, we will denote the characteristic vector $w^h$ of the headline $h$ ($h = 1, 2, ..., S_j$) as follows:

$$w^h = (w^h_{t_1}, w^h_{t_2}, ..., w^h_{t_m}), \qquad (1)$$

where $m$ is the number of different terms in the headline $h$ and $t_k$ denotes each term.

Using the scheme *tf*, or term frequency, each element $w^h_{t_k}$ of $w^h$ is defined as follows:

$$w^h_{t_k} = \frac{tf_{h,k}}{\sum_{s=1}^{m} tf_{h,s}}, \qquad (2)$$

where $tf_{h,k}$ is the frequency of the term $t_k$ in the headline $h$.

Then, we define $P_s$ as:

$$P_s = (ps_{t_1}, ps_{t_2}, ..., ps_{t_u}), \qquad (3)$$

where $u$ is the number of different terms in all headlines chosen in the session $j$ and $t_k$ denotes each term.

And we define each element $ps_{t_k}$ using the equation (2) as follows:

$$ps_{t_k} = \frac{1}{S_j} \sum_{h=1}^{S_j} w^h_{t_k}, \qquad (4)$$

The user profile *P* will be denoted also by a vector:

$$P = (p_{t_1}, p_{t_2}, ..., p_{t_n}), \qquad (5)$$

where *n* is the number of different terms in the profile *P* and $t_k$ denotes each term.

Each element $p_{t_k}$ is defined in turn as follows:

$$p_{t_k} = \frac{1}{S_j} \sum_{h=1}^{S_j} w_{t_k}^h, \qquad \text{for } j = 1 \qquad (6)$$

$$p_{t_k} = a \cdot p_{t_k} + b \cdot \frac{1}{S_j} \sum_{h=1}^{S_j} w_{t_k}^h, \qquad \text{for } j > 1 \qquad (7)$$

where *a* and *b* are two constants that satisfy *a+b=1*, whose values are experimentally established, and $S_j$ is the number of headlines which have been chosen by the user in the session *j*. The constant *a* indicates the relative relevance that is given to the user's preferences stored in the profile *P* and the constant *b* indicates the relative relevance assigned to the user's preferences detected in the session *j*.

Also it is considered a characteristic vector $w^{hr}$ composed of the terms that appear in the summary *r* associated with a headline *h*.

Thus, we define $Sr_j$ *(j= 1, 2,..., R)* as the number of headlines with associated summary which have been chosen by the user in the session *j*. For each session, will be elaborated a profile $P_r$ with the terms of the summaries applying the following process. First of all, we will denote the characteristic vector $w^{hr}$ of the summary associated with a headline *h* (*h = 1, 2,..., $Sr_j$*) as:

$$w^{hr} = (w_{t_1}^{hr}, w_{t_2}^{hr}, ..., w_{t_v}^{hr}), \qquad (8)$$

where *v* is the number of different terms in the summary *r* associated with headline *h* and $t_k$ denotes each term. Using the *tf* scheme of the frequency of the term, each element $w_{t_k}^{hr}$ in $w^{hr}$ is defined as follows:

$$w_{t_k}^{hr} = \frac{tf_{hr,k}}{\sum_{s=1}^{v} tf_{hr,s}}, \qquad (9)$$

where $tf_{hr,k}$ is the frequency of the term $t_k$ in the summary *r* associated with the headline *h*.

Then, we define $P_r$ as:

$$P_r = (pr_{t_1}, pr_{t_2}, ..., pr_{t_z}),  \qquad (10)$$

where $z$ is the number of different terms in all the summaries chosen in the session $j$ and $t_k$ denotes each term.

We define each element $pr_{t_k}$ using the formulation (9) as follows:

$$pr_{t_k} = \frac{1}{Sr_j} \sum_{h=1}^{Sr_j} w_{t_k}^{hr} \qquad (11)$$

The **elaboration of the profile of user** $P$ at the end of each session proceeds as follows: let $P_j$ be the user profile stored after the session $j$, and let be $P_{s,\,j+1}$ the profile of the session $j+1$. Then, for all $t_k \in (t_1, t_2, ..., t_u)$, where $u$ is the number of different terms found in the session $j+1$ and $t_k$ denotes each term, the profile $P_{j+1}$ built at the end of session $j+1$ is given by the following expressions:

$$P_{j+1} = (0.5 \cdot P_j + 0.5 \cdot P_{s,\,j+1}) + P_{r,\,j+1} \qquad \text{if } p_{t_k} \in P_j \qquad (12)$$

$$P_{j+1} = P_{s,\,j+1} + P_{r,\,j+1} \qquad \text{if } p_{t_k} \notin P_j \qquad (13)$$

where $P_{r,\,j+1}$ is the profile $P_r$ in the session $j+1$.

### 3.2. Computing headline scores

In order to compute the score associated with a headline $h$, we will compare its corresponding characteristic vector $w^h = (w_{t_1}^h, w_{t_2}^h, ..., w_{t_m}^h)$ with the user profile $P = (p_{t_1}, p_{t_2}, ..., p_{t_n})$.

The similarity, $sim(P, w^h)$, between the user profile $P$ and the characteristic vector of the headline $h$, $w^h$, is calculated applying the **cosine measure**[8]:

$$sim(P, w^h) = \frac{P \cdot w^h}{|P| \cdot |w^h|} = \frac{\sum_{k=1}^{m} p_{t_k} \cdot w_{t_k}^h}{\sqrt{\sum_{k=1}^{m}(p_{t_k})^2 \cdot \sum_{k=1}^{m}(w_{t_k}^h)^2}} \qquad (14)$$

The similarity value given by equation (14) is the score for headline $h$ according to the user profile $P$. Then the news headlines are ordered for each user according to his/her profile, presenting in the first positions those headlines with greater score.

Another way of computing the score associated with a headline $h$ consists in applying a **simple criterion of binary relevancy**: a document is relevant, or not, if it contains the requested word, without discriminating between different degrees of relevance, that is to say, using a boolean algorithm. This simple criterion of binary relevancy is introduced to be compared with algorithm NectaRSS afterwards.

Thus, given the corresponding characteristic vector of the headline $h$, $w^h = (w^h_{t_1}, w^h_{t_2}, ..., w^h_{t_m})$, where $m$ is the number of different terms and $t_k$ denotes each term, and the user profile $P = (p_{t_1}, p_{t_2}, ..., p_{t_n})$, where $n$ is the number of different terms, then the similarity, $sim(P, w^h)$, between the user profile $P$ and the characteristic vector of the headline $h$, $w^h$, will be given by the following expressions:

$$sim(P, w^h) = 0 \qquad \text{if } p_{t_k} \notin P \qquad \forall t_k \in (t_1, t_2, ..., t_m) \qquad (15)$$

$$sim(P, w^h) = 1 \qquad \text{in other case} \qquad (16)$$

The similarity value given by the expressions (15) and (16) is the score of the headline $h$ according to the user profile $P$. It is enough that any term of the headline is found in the user profile to consider that there exists similarity between both of them and assign a score value of 1 to this similarity. Then the news headlines are sorted for the user showing first those with score 1.

## *4. Experiments and results*

In order to contrast and determine the validity of results, several sessions were carried out with different real users. The user was offered a headlines list, ordered by score. Then, he/she had to choose those headlines of his/her interest. The number of headlines offered permitted the user to see them all without the need of vertical page displacements. 15 users tested the system taking into account that their thematic interests should be heterogeneous. At the beginning of each experiment, the user profile was empty and during the sessions it was elaborated and completed.

### 4.1 Measures for the experimental evaluation of the system

The following measures have been used to check the validity of the proposed method.

**Average score of a set of headlines and maximum mean score**

In each session the user is offered a certain number T of headlines and he/she must choose those of his/her interest, termed the *chosen headlines* or $E(T)$. Then, the average score or $\overline{p(E(T))}$ is calculated for the set of headlines selected by the user in that session. On the other hand, it can be calculated a maximum average score value or $\overline{p_{max}(T)}$ for the set of headlines. It is obtained when the news ($N$) chosen are the same to the first $N$ headlines (in score order) offered by the system in a given session. To quantify the relationship between the value $\overline{p(E(T))}$ of the headlines selected by the user and the value $\overline{p_{max}(T)}$, the **rate** $C_D$ is defined:

$$C_D = \frac{\overline{p(E(T))}}{\overline{p_{max}(T)}}, \qquad (17)$$

where $\overline{p_{max}(T)}$ is the average of the first *N* score values associated with the *N* headlines with greater score among those offered to the user, where *N* is the number of headlines chosen by the user.

**The R-Precision**

According to Baeza [1], a simple summary value for a set of headlines offered in score order will be generated. For this, it will be calculated the *precision* in the position *R* of the order, being *R* the total number of relevant headlines of the session, that in the case of NectaRSS is the number of headlines which the user has chosen among those offered by the system.

The calculation of the R-Precision is then defined as:

$$RP(i) = \frac{posR(E(T_i))}{card(E(T_i))}, \qquad (18)$$

in which *posR(E(T_i))* is the number of headlines chosen among the first R headlines orderly offered to the user in the session *i*, and the value of *card(E(T_i))* is the total number of headlines chosen in such session.

## 4.2 Test of the Algorithm NectaRSS with different users

The NectaRSS has been tested with different users. In each session the user is shown a selection of 14 headlines ordered by score; this quantity has been chosen so that all headlines are shown at once in a single page, without forcing the user to scroll down to get them.

Every one of the 15 voluntary users carried on 2 training sessions and 30 experimental sessions, choosing the information of their interest from among those 14 headlines offered by the system; experimental sessions are also used to improve the user's profile, at the same time that it is showing ranked headlines (ORDER experiment). Furthermore, in order to compare results, the participants run over 30 new test sessions where every user chooses at random headlines of his/her interest among 14 offered (RANDOM sub-experiment). Obviously, since there is no user profile at the beginning of the experiment, headlines in the first training session are randomly ordered.

Results obtained for $C_D$ in the $30^{th}$ experimental session for the 15 users are shown in Figure 1.

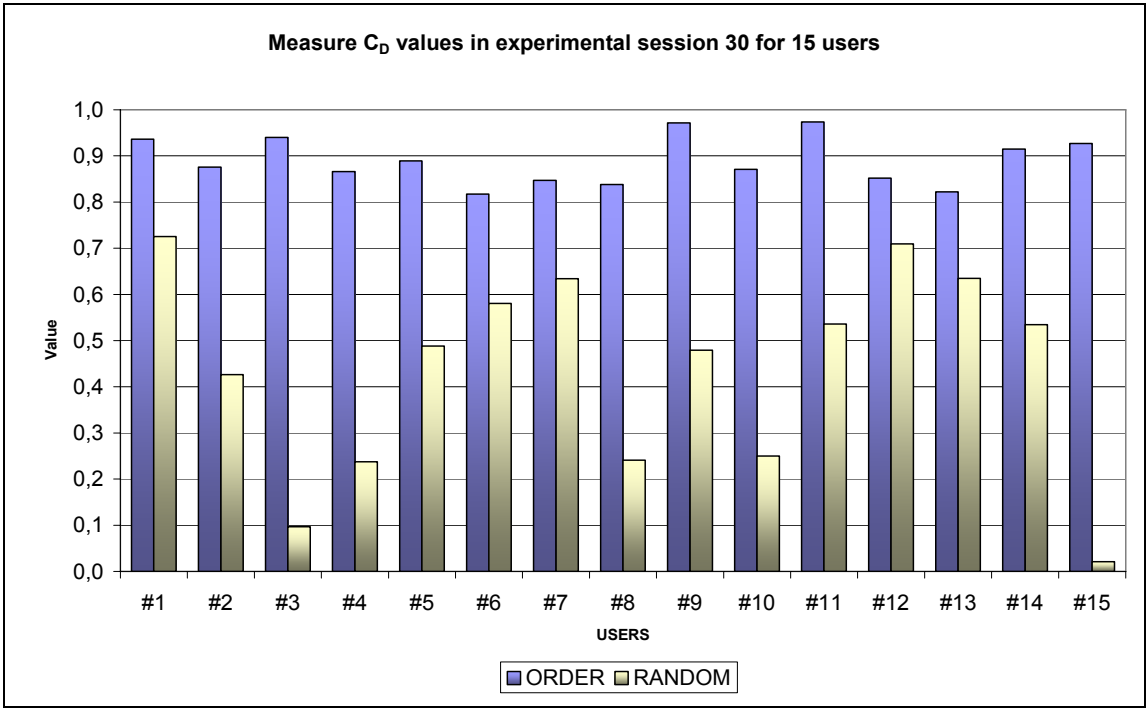

Figure 1. Results obtained by 15 users for the rate $C_D$ in the experimental session 30, when headlines are offered ranked by the NectaRSS algorithm (ORDER case), and at random (RANDOM case). As can be seen, NectaRSS outperforms the case where headlines are offered at random.

As can be seen, all users yield better values in the ORDER case, than in the RANDOM case. This means that the headlines chosen by the user in the ORDER case have greater score than those chosen in the RANDOM case, that is, the user finds a larger amount of interesting headlines among these presented to him/her when the user profile computed by NectaRSS is used to rank headlines.

The R-Precision measure has only been applied to the ORDER case, since it needs an ordered set of headlines to compute the precision for position R in the ranked headlines. To compare the R-Precision throughout 30 experimental sessions the user with the worst (#8) and the best (#11) average for this measure, have been chosen, which act as *de facto* upper and lower bound in performance; the rest of users will yield figures between these two. Figure 2 shows graphically the values of the R-Precision obtained by these users in 30 experimental sessions along with the trend line from each one, Linear (User # 8) and Linear (User #11).

As it can be seen in the figure below, the slope of the trend line is positive in both cases. This indicates that the system improves its headline rankings with the number of sessions, which is a desired behaviour, and it shows that the ranking offered by RSS increasingly matches the one that would have been done by the user himself.

This means that the user profile constructed by NectaRSS really characterizes the corresponding user and allows improving the response of the information retrieval systems consulted by this user.

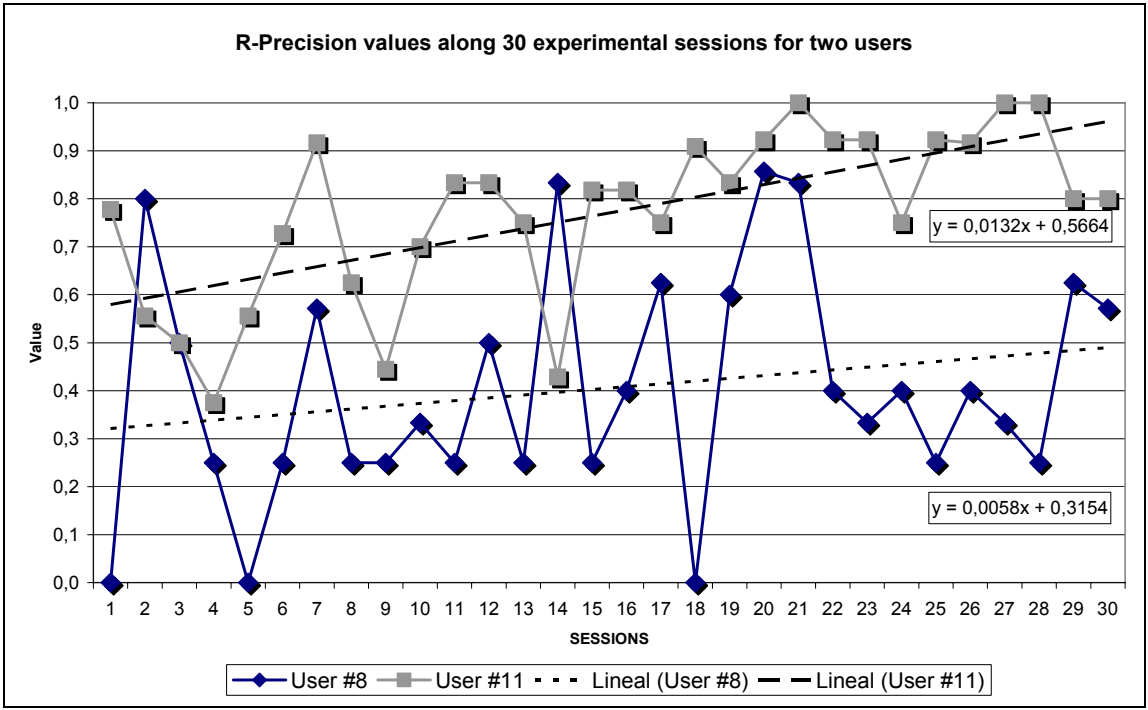

**Figure 2. Results obtained by the user #8 and by the user #11 for the R-Precision throughout 30 experimental sessions, together with the trend lines from the data. It is observed a favourable evolution of the R-Precision.**

## 4.3 Headline scoring using a boolean algorithm

We have also tried to prove that NectaRSS outperforms more naïve algorithms, and, in particular, that the vector space representation is better than a more straightforward purely boolean (binary) representation, which has been introduced in section 3.2.

In this experiment the 15 voluntary users submitted to 30 additional experimental sessions with NectaRSS, configured now to score incoming newsitems through a simple criterion of binary relevancy.

In the new sessions, the users have been presented the same set of news that was used to score the information with the cosine measure. This allows us to compare the results obtained with NectaRSS, in the ORDER case, with those obtained using the boolean algorithm.

The average $C_D$ for the 15 users can be seen in figure 3, which shows the $C_D$ rate by user for the two cases: the cosine measure and the boolean algorithm. Both cases are also considered for the R-Precision; results are shown in figure 4.

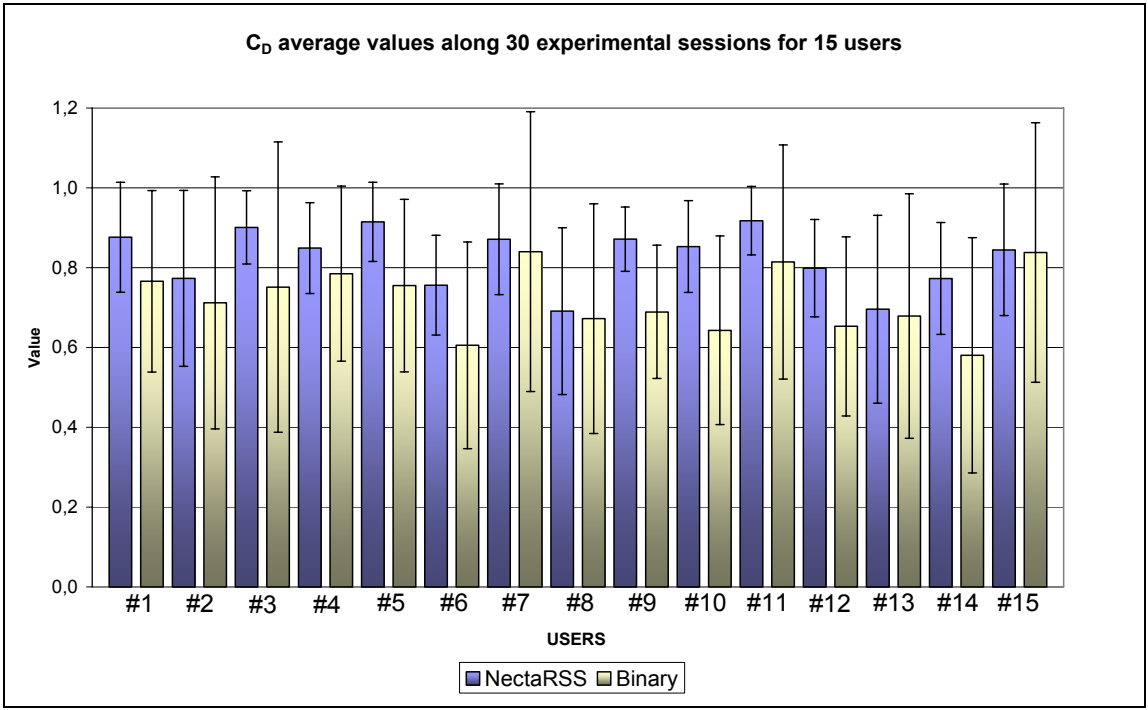

**Figure 3.** $C_D$ average measures along 30 experimental sessions, using the cosine measure to score the headlines (NectaRSS) and using a binary algorithm to calculate such score (Binary).

As it can be seen above, the $C_D$ average rates along the 30 sessions is better for all users in the NectaRSS case, than in the Binary case. We can conclude that the headlines presented to the user applying the binary algorithm are much less related to the user's interest, than the ones presented by applying NectaRSS. The R-Precision measure for both kinds of distance measures have also been compared, and results are shown in Figure 4. In this case also the improvement in rankings shown by NectaRSS is better for the 15 users than the one shown by the binary measure.

In order to have a global idea of the behaviour of each considered algorithm, we have analyzed the results of both, the R-precision measure and the $C_D$ rate, obtained by all the users at each experimental session. At each session, we compute the difference between the average values, for the 15 users, obtained with the NectaRSS algorithm (average NectaRSS), and the binary algorithm (average binary). Positive values indicate that NectaRSS beats the binary algorithm.

Figure 5 and figure 6 shows the values of this difference for the $C_D$ rate and the R-precision measure respectively, along the 30 experimental sessions.

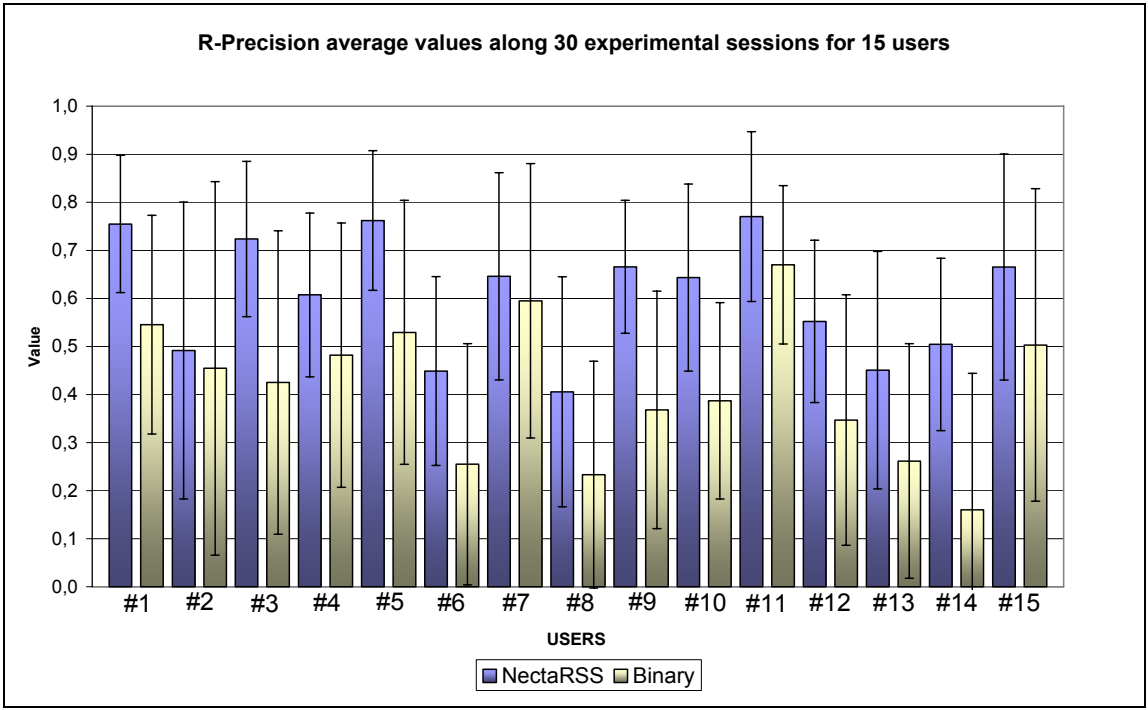

**Figure 4. R-Precision average measure along of 30 experimental sessions, when the cosine measure is used to score the headlines (NectaRSS) and using a binary algorithm to compute such score (Binary).**

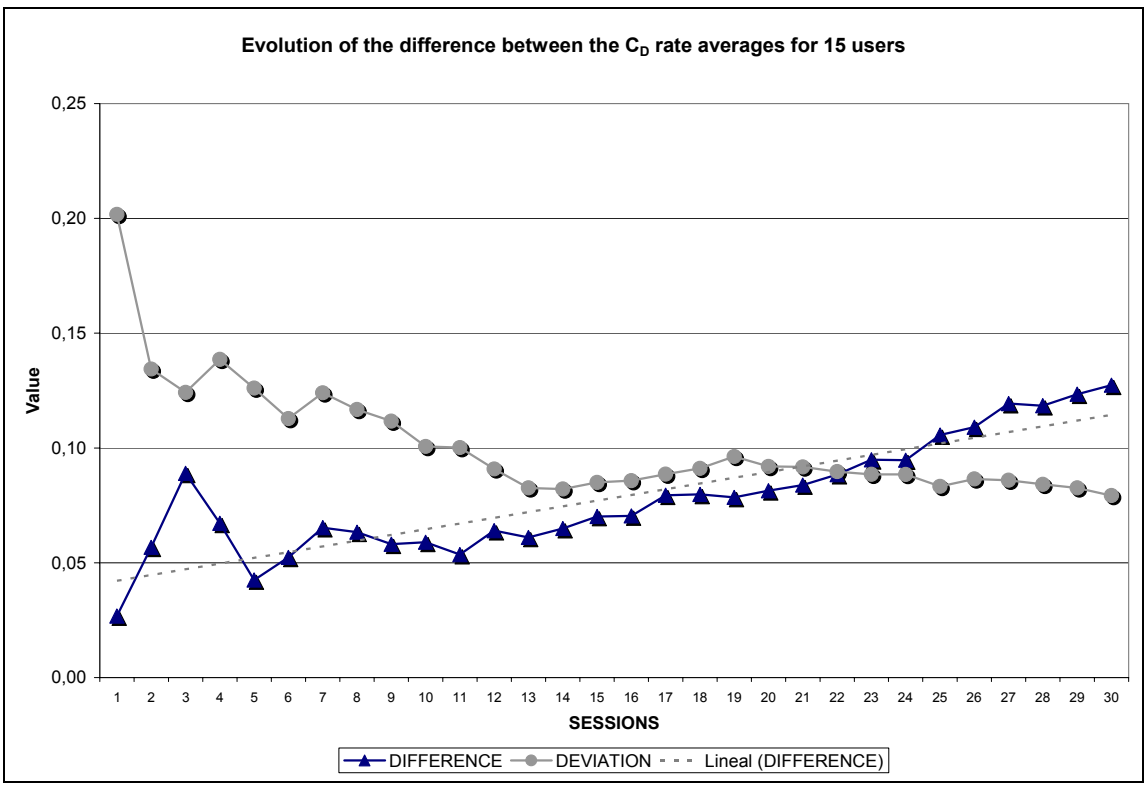

**Figure 5. Evolution of the difference between averages (with standard deviation) with respect to the $C_D$ rate during 30 experimental sessions.**

Figure 5 shows that the difference between the $C_D$ rate averages increases along the sessions, what means that the advantage of NectaRSS over the binary algorithm gets larger with the system training. We can also observe that the deviation clearly decreases with the sessions, denoting that the values are less scattered. Thus, for all the users goes observing a progressive improvement of the algorithm NectaRSS respect to the purely binary algorithm.

Figure 6 shows the difference between the R-Precision values. This difference also increases with the sessions, indicating that the advantage of NectaRSS over the binary algorithm improves with the sessions. The deviation also decreases with the sessions.

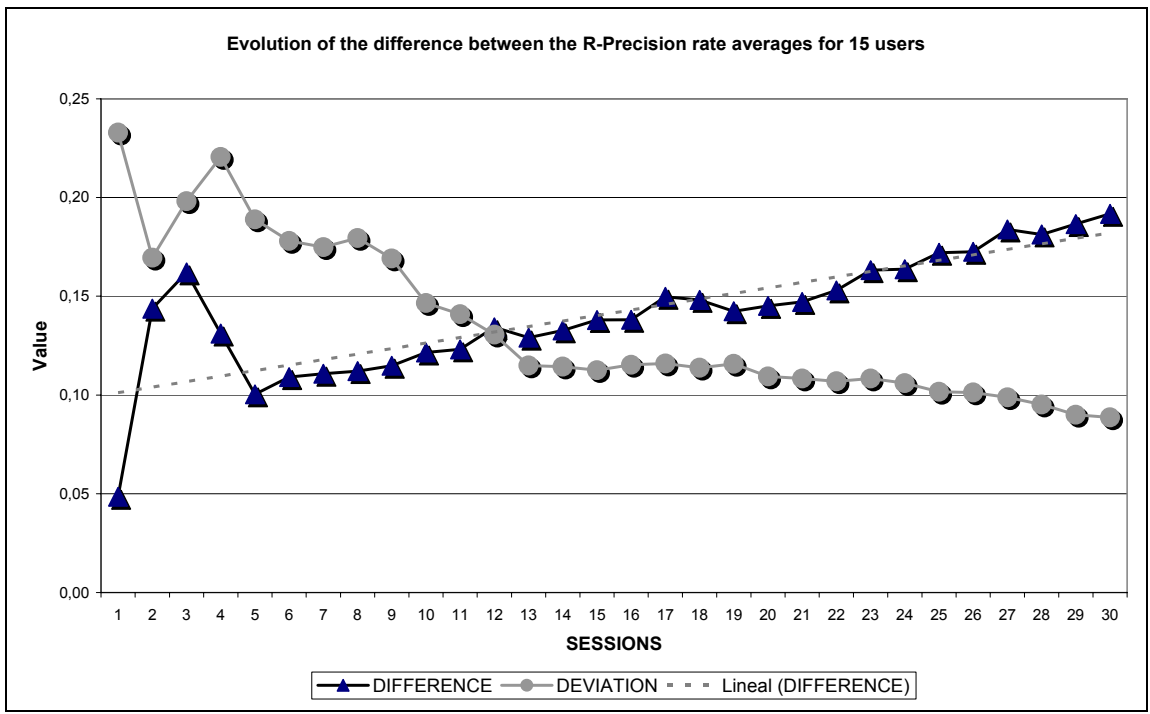

**Figure 6. Evolution of the difference between averages (with standard deviation) with respect to the R-Precision rate during 30 experimental sessions.**

## 5. CONCLUSIONS AND DISCUSSION

Considering the above results, we can assert that the newsitems scoring achieved applying the user profile computed via the NectaRSS algorithm is significantly useful. The user is shown more interesting documents or, at least, more according to his/her preferences. These advantages of the proposed system have been shown for different and heterogeneous users (in the sense that their technical background and preferences are different). Comparing the NectaRSS algorithm with a different way of scoring the information retrieved, namely a simple binary algorithm, it is observed that in the first case the ranking of the information is more appropriate, and moreover, the improvement of the system response is faster.

Furthermore it has been observed a positive evolution of the algorithm NectaRSS throughout the experimental sessions for all the users, and so on better than simple binary algorithm and with values progressively less discontinuous.

We can conclude that our system NectaRSS **has been able to endow a certain degree of "intelligence" to a typical content aggregator**, filtering its RSS contents better than either a random system or a system with simple binary scoring. This approach is novel in two different senses: first, the profile building algorithm has been designed *ab initio*, although it is based on mainstream information retrieval ideas, and second, it is the first time this kind of algorithms has been used on feed aggregators.

The proposed system can be improved along two different lines of future work:

- Application of linguistic analysis to the retrieved information, which allows the fine tuning of the features used to characterize the documents by selecting particular types of words, such as names, or using stemming over extracted words.

- Use of web text collections[5] specifically defined for the evaluation of retrieval information systems. Since queries and relevance assessments are available for these collections, we plan to use them to evaluate our system.

- More extensive experimentation, using more users and a wider source selection, to confirm these data and analyze if there are differences among users.

## *ACKNOWLEDGEMENTS*

This paper has been funded in part by project TIC2003-09481-C04-04, of the Spanish Ministry of Science and Technology, and a project awarded the Quality and Innovation department of the University of Granada, and by Resolution 8-7-2004 of General Management of Educative Innovation and Professorship Formation of the Science Department of the Regional Government.## *References*

---

[5] *Web Research Collections. TREC Web & Terabyte Tracks.*
*http://ir.dcs.gla.ac.uk/test_collections/*